\documentclass{article}

\usepackage{PRIMEarxiv}
\usepackage[utf8]{inputenc} 
\usepackage[T1]{fontenc}    
\usepackage{hyperref}       
\usepackage{url}            
\usepackage{booktabs}       
\usepackage{amsfonts}       
\usepackage{nicefrac}       
\usepackage{microtype}      
\usepackage{cite}
\usepackage{titlesec}
\usepackage{fancyhdr}       
\usepackage{graphicx}       
\graphicspath{{media/}}     
\usepackage{caption}
\usepackage{subcaption}
\usepackage{amsmath}   
\usepackage{float}     
\usepackage{algorithm}
\usepackage{algorithmic}
\usepackage{indentfirst}
\usepackage{multirow}
\usepackage{eqlist}

\newcounter{alphsection}
\newcommand{\alphsection}[1]{%
  \refstepcounter{alphsection}%
  \subsection*{\alph{alphsection}.~#1}%
  \addcontentsline{toc}{subsection}{\alph{alphsection}.~#1}%
}

\title{Using Physics Informed Generative Adversarial Networks to Model 3D porous media

}

\author{
  Zihan Ren \\
  Department of Energy and Mineral Engineering \\
  Pennsylvania State University \\
  State College\\
  \texttt{zur74@psu.edu} \\
   \And
  Sanjay Srinivasan \\
  Department of Energy and Mineral Engineering \\
  Pennsylvania State University \\
  State College\\
  \texttt{szs27@psu.edu} \\
}

\begin{document}
\maketitle

\begin{abstract}



Micro-CT scanning of rocks significantly enhances our understanding of pore-scale physics in porous media. With advancements in pore-scale simulation methods, such as pore network models and the Lattice Boltzmann method, it is now possible to accurately simulate multiphase flow properties, including relative permeability, from CT-scanned rock samples. These physical properties are crucial for describing the multiphase flow behavior of CO$_2$-brine systems during CCUS CO$_2$ storage. However, the limited number of CT-scanned samples and the difficulty in linking the pore-networks to field-scale rock properties often renders it difficult to use pore-scale simulated properties in realistic field-scale reservoir simulations. Deep learning-driven approaches to construct synthetic 3D rock microstructures make it possible to simulate variability in CT rock structures, which can be subsequently used to compute representative rock properties and flow functions. Nonetheless, most current deep learning-based 3D rock structure synthesis is unconstrained by any rock properties that may be derived from well observations, thereby lacking a direct link between 3D pore-scale structures and field-scale observations. We present a method to construct 3D rock structures constrained to observed rock properties using generative adversarial networks (GANs) with conditioning accomplished through a gradual Gaussian deformation process.

We begin by pre-training a Wasserstein GAN to reconstruct 3D rock structures. Subsequently, we use a pore network model simulator to compute rock properties. The latent vectors for image generation in GAN are progressively altered using the Gaussian deformation approach to produce 3D rock structures constrained by well-derived conditioning data. This GAN and Gaussian deformation approach enables high-resolution synthetic image generation and reproduces user-defined rock properties such as porosity, permeability, and pore size distribution. Our research provides a novel way to link GAN-generated models to field-derived quantities, offering a significant step towards designing a systematic machine-learning workflow that interpolates subsurface properties by combining both pore-scale and field-scale data. These are important steps towards the upscaling of crucial multiphase physical properties such as relative permeability or capillary pressure from pore scale to field scale.
\end{abstract}


\section{Introduction} 

\label{sec:intro}
The analysis and reconstruction of 3D micro-CT porous media is crucial for numerous engineering applications, particularly in digital rock analysis and material science applications. For instance, in geologic carbon sequestration (GCS), understanding the multiphase interaction between CO$_2$ and brine is vital for predicting flow behavior of the gas plume and the long-term storage potential of the reservoir. Traditionally, the acquisition of physical properties such as relative permeability has relied on laboratory measurements or physical simulations on 3D CT images. While laboratory measurements are generally accurate, they are labor-intensive, time-consuming, and often represent only a specific rock type found in the subsurface~\cite{2020Zhao_krpred}. In contrast, direct pore-scale simulations on 3D micro-CT images offer a more flexible and varied approach, allowing for the manipulation of different physical characteristics (e.g., capillary number, wettability) to better observe the sensitivity to different physical factors~\cite{2021Prakash_krEoS}. 

According to results obtained using micromodels (miniaturized artificial pore network models) \cite{2016Senyou_psonkrk, 2014Xu_poregeoonkr} and from laboratory experiments \cite{2013Zhang_krheterogeneity}, pore structure properties such as pore throat ratio, coordination number, shape factors, and pore throat orientation, as well as porosity have a significant impact on relative and absolute permeability. Moreover, the spatial distribution of porosity will influence the macro scale characteristics of relative and absolute permeability. Representations of porous media using PNMs lack the authenticity to represent property variations in real subsurface formations and thus cannot be used for practical applications. One potential solution is to develop models of porous media that reflect realistic porosity or permeability distribution so that multiphase flow properties simulated using such a medium can be utilized at field scale more consistently. However, micro-CT measurements are typically sparse and underrepresent features at larger scales, such as field-scale observations. 

This limitation has driven research efforts towards developing models for synthetic porous media that can represent pore features accurately and have the ability to model increased variability in pore characteristics, with the ultimate goal of improving our ability to interpolate and upscale multiphase flow process properties while honoring data at different scales. Statistical-based reconstruction such as using multiple-point-statistics-based method~(MPS) is able to generate certain geology realizations using training image template that match with production response or post simulation results~\cite{2010Sanjay_multipoint_reservoir,2014JefCaers_MPS} and has been applied for porous media reconstruction \cite{2005_pnmreconstruct_blunt, 2013Sahimi_2pointreconstruct} with varying degree of success. Recent advancements in deep learning offer potentially more comprehensive solutions to reconstruct complex target distributions while assimilating data from different sources by using deep generative neural networks. Some successful examples include using AttentionGAN in image synthesis \cite{2017TaoXu_AttnGAN}, 2D/3D image rendering \cite{2015_vaeDCIGETejas}, scene reconstruction \cite{2020Nerf_Ben}. These approaches have primarily focused on building generative models conditioned to semantic meanings, especially prompts, and tokens. Compared with MPS, deep learning-based generative models have a faster reconstruction speed once training has been completed and may be more capable of capturing complex random functions from training data\cite{2017MosserGAN}, because they have a larger parameter space.

Generative Adversarial Networks (GANs) are two-network systems that learn data representation through an adversarial training process \cite{2014Goodfellow_GAN}. During the training process, the generator is used to synthesize images while the discriminator's objective is to distinguish the difference between the synthesized image and the real image. This adversarial training process significantly improves image quality compared to other generative models, such as Variational Autoencoders (VAEs) \cite{2021Sam_compareGANVAE_other}. In terms of image reconstruction, GANs can be combined with CNNs to reconstruct complex image features. For instance, in subsurface imaging, Deep Convolutional GAN (DCGAN) has been used to reconstruct 3D MicroCT images of rocks \cite{2017MosserGAN, 2018MosserGAN}. However, training of GANs is notoriously unstable \cite{2017WGAN_Arjovsky}. To address this, various GAN variants have been developed in recent years, featuring different neural network structures and loss functions to stabilize training and evaluation. For example, by replacing the binary entropy loss function of the original GAN with a Wasserstein distance based loss function   the resultant Wasserstein GAN with gradient penalty (WGAN-GP) significantly improves and stabilizes the GAN training process \cite{2017WGAN_Arjovsky, 2017WGANGP_Ishaan}. Moreover, style is injected in the form of progressive training strategies \cite{2017ProgressiveGAN}, to produce high-fidelity, high-resolution images, such as synthetic human faces \cite{2018StyleGAN_Tero, 2019Tero_stylegan2} using StyleGAN.\par

Utilizing a GAN-based deep learning approach to reconstruct porous media offers several advantages. Several GAN-based applications have demonstrated promising results for capturing complex 2D and 3D subsurface spatial relationships. For instance, GANs can be employed to reconstruct high-resolution channel facies \cite{2020ChanGANsubchannel} while accounting for varying spatial proportions of channels. In the realm of digital rocks, DCGAN has been utilized to reconstruct 2D and 3D micro-CT images of Berea sandstone, bead packs, and oolitic Ketton limestone \cite{2017MosserGAN, 2018MosserGAN, 2021AssembleGAN_Sung}, while honoring petrophysical and Minkowski functional statistical distributions from training samples. While GAN-based approaches demonstrate high-fidelity reconstruction capabilities, the challenge of constraining these methods to honor physical conditional data from diverse sources remains an active area of research.\par

There are generally two ways to control GAN generation. The first involves embedding conditional vectors in the training image space and latent vector space ($z$ ), as originally implemented by the conditional GAN algorithm \cite{2014CGAN_Mehdi}. In subsurface modeling, similar ideas have been applied to develop earth models conditioned to 3D sparse rock facies data using GANSim-3D \cite{2020SongGANSIM, 2022Song_GANSim3D}, and to reconstruct 3D micro-CT images from 2D slices \cite{2020Coiffier_2d3dusingGANCT, 2021GAN_2D3D_stevekench}. Another conditional generation approach involves manipulating the latent space ( $z$ ) directly \cite{2016Alec_DCGANlatentspace, 2019Shen_latentspaceGAN} to find the target structure. By establishing a mapping function between the latent space ( $z$) and physical properties, different realizations of 3D Micro-CT images can be generated, constrained by observation properties. Building on this principle, Markov Chain Monte Carlo (MCMC) sampling algorithms have been utilized to search for appropriate latent vectors for conditionally reconstructing specific geostatistical realizations within an inversion algorithm \cite{2018Laloy_GANtrainingimage}. However, MCMC is computationally expensive \cite{2022Song_GANSim3D} for finding the target latent vector. One potential solution is to train another ML model to parameterize the mapping between latent space and target physical properties \cite{2019chan_ganparameteric}. These kinds of approaches generally require a second stage of training and predefined physical attributes for efficiently performing the training.\par

\begin{figure}[hbt]
    \centering
    \includegraphics[width = \textwidth]{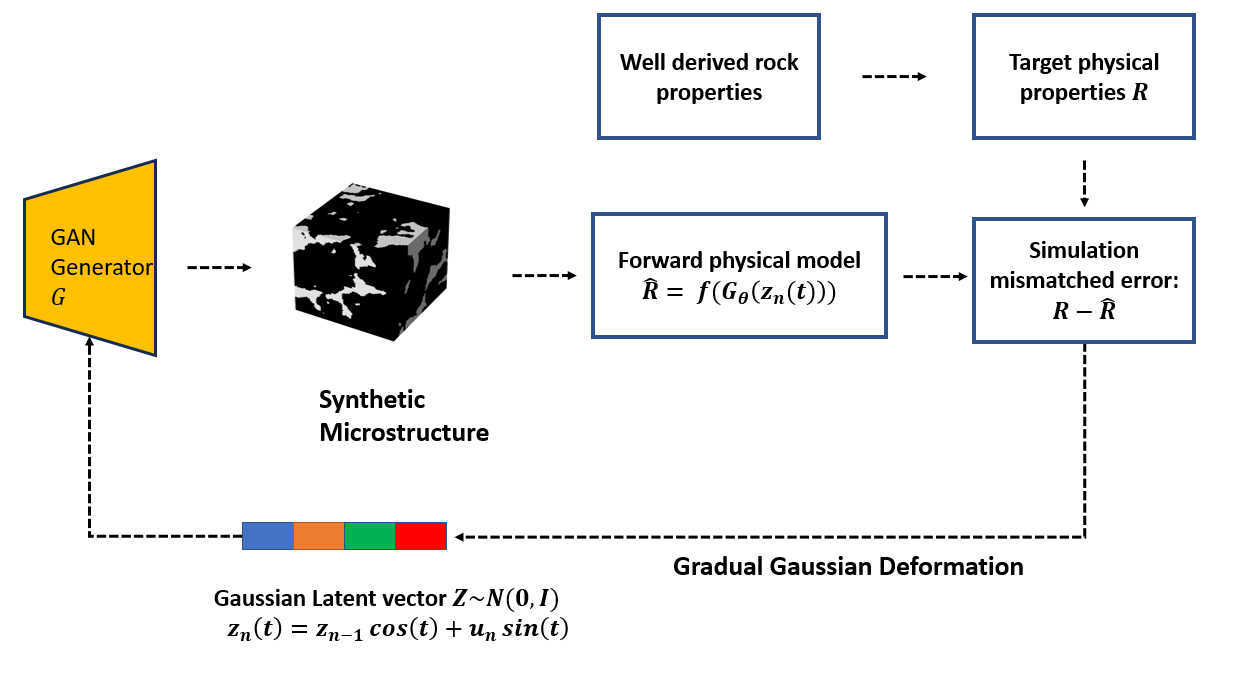}
    \caption{Workflow of controllable generation using GAN and physics informed Gradual Gaussian Deformation
}
    \label{method_fig:GAN_workflow}
\end{figure}\par

Inspired by ideas of working on latent space for conditioning the GAN models, approaches to concatenate data generated by physical simulation directly into the latent space within an inversion framework has become more popular recently. By integrating a differentiable physical simulator on top of GAN-generated subsurface realizations, the mismatch error between the GAN-derived physical responses and the observed data can be directly backpropagated to the generator and into the latent vector $z$. Building on this idea, GANs are used to derive forward and inverse solutions of partial differential equations~\cite{2021CGANPDE_kad} or to construct hybrid physics-based and data-driven models for solving inverse problems by manipulating the latent space \cite{2022wu_vaeinverse}. More recently, a GAN and actor critic reinforcement learning framework~(GAN-AC) was developed to search for stochastic parameters to control user-defined GAN generation \cite{2022_gan_RL}. The reinforcement learning agent receives feedback from a physical simulator to gradually calibrate the model using injected random noise that have been used to fine tune the reconstruction process.

There are some limitations facing current approaches to condition GAN models to data. The conditional GAN~(cGAN) workflow is straightforward but often not feasible for reconstructing multivariate physically-driven realizations, especially when involving multiple, inter-correlated physical parameters. However, the cGAN-based approach necessitates labeling conditioning data for each training image, which is both computationally expensive and render the approach inflexible in real practice due to the need to pre-define conditional variables during GAN training. In contrast, manipulating the GAN's latent space or stochastic parameters directly during the post-training process offers greater flexibility for reconstructing 3D objects. However, the challenge remains in identifying and searching for these parameters. The GAN-AC framework that utilizes a reinforcement learning feedback loop can perform an effective searching process. However, that necessitates a post-training process and a fixed physical simulator to train the reinforcement learning agent. However, most commercial physical simulation software are non-differentiable, rendering it hard to incorporate into a differentiable inversion framework such as in \cite{2022wu_vaeinverse}. Given the existence of multiple physical simulation approaches, such as Pore Network Modeling (PNM) or the Lattice Boltzmann Method, each capable of incorporating varied physical factors like wettability and capillary number, a more flexible and controllable generation framework is necessary to work harmoniously with physical simulation software. This approach should efficiently optimize the latent vector in GAN without extensive dependence on a fixed post-training environment, thereby balancing accuracy with flexibility.\par

To address the above challenges, we propose a two-stage approach to generate 3D micro-structure of sandstone, conditioned to target rock properties. Initially, we pretrain a Wasserstein GAN with gradient penalty (WGAN-GP) to generate 3D $\mu CT$ images. In the subsequent stage, we employ a gradual Gaussian perturbation approach combined with a physical simulator such as Pore Network Modeling (PNM) to directly optimize the Gaussian-based latent space $z$. This optimization ensures that the generated rock structures align with physical simulation outputs. Our method guarantees that the optimized latent vectors can produce 3D porous media consistent with user-defined rock properties. This approach effectively controls GAN generation to preserve critical pore structure parameters, such as porosity $\phi$, permeability $k$, mean pore and throat size diameter during the generation process. Porosity and permeability are common attributes that can be derived at well or field scale, thus our approach serves as a fundamental stepping-stone to upscaling multiphase transport properties from pore-scale to field scale simulation.\par

\section{Methods}
\label{sec:Method}

\subsection{Generative Adversarial Networks}
\label{subsec:method/GAN}
GAN consists of a discriminator ($D$) and a generator ($G$). The goal of the discriminator is to evaluate and distinguish between generated samples $\hat{y}$ (fake) and real training samples $y$ (real). The generator aims to generate samples $\hat{y}$, decoded from a latent vector $z \sim p_z(z)$, so that $\hat{y}$ can deceive the discriminator into incorrectly classifying fake generated images $\hat{y}$ as real. This min-max competition between $G$ and $D$ is mathematically characterized by the adversarial loss \cite{2014Goodfellow_GAN} described in equation~\ref{methodeq:ganloss}, where $E$ denotes the expectation (average) operator over the distributions $y \sim p(y)$, representing the training image space. Parameters $\theta$ and $\omega$ represent the generator's and discriminator's parameter spaces, respectively.\par

The discriminator $D_\omega$ aims to minimize the binary cross-entropy loss on its predictions, as described in Equation~\ref{methodeq:Dloss}. A lower binary cross-entropy loss indicates a more confident discriminator that can effectively distinguish real samples from fake ones. Conversely, the generator training objective is equivalent to minimizing the loss function described in equation~\ref{methodeq:Gloss}, which is akin to maximizing the probability that the discriminator classifies fake samples $\hat{y}$ as real. Theoretically, the optimal stopping criterion for the GAN training process is reached when the generator loss and discriminator loss are approximately equal, a state known as Nash equilibrium \cite{2016Ian:deeplearning}. However, in practice, training GANs to this point can be challenging and often relies on other metrics, such as reconstruction quality, to evaluate progress.

\begin{equation}\label{methodeq:ganloss}
\min_\theta \max_\omega \mathcal{L}(D_{\omega},G_{\theta}) = \mathbb{E}_{y \sim p(y)}[\log(D_{\omega}(y))] + \mathbb{E}_{z \sim p_z(z)}[\log(1-D_{\omega}(G_{\theta}(z)))]
\end{equation}\par

\begin{equation}\label{methodeq:Dloss}
\mathcal{L}(D_{\omega}) = - \left[ \mathbb{E}_{y \sim p(y)}[\log(D_{\omega}(y))] + \mathbb{E}_{z \sim p_z(z)}[\log(1 - D_{\omega}(G_{\theta}(z)))] \right]
\end{equation}

\begin{equation}\label{methodeq:Gloss}
\mathcal{L}(G_{\theta}) =  -\mathbb{E}_{z \sim p_z(z)}[\log(D_{\omega}(G_{\theta}(z)))]
\end{equation}

This adversarial training scheme can be customized for different applications. In the context of image generation, the Deep Convolutional Generative Adversarial Networks (DCGAN) architecture is often utilized, where both the generator and discriminator are Convolutional Neural Networks (CNNs), which are well-suited for processing grid-based data such as images. The generator aims to upsample a Gaussian-based latent vector $z$ to obtain synthetic images, whereas the discriminator aims to downsample the training image into a critic score. In the generator $G_\theta$, the Gaussian vector $z$ is non-linearly transformed into a synthetic image $\hat{y}$ to deceive the discriminator into misclassifying the synthetic image as real.

The original GAN, which utilized binary cross-entropy loss, often suffers from instability and mode collapse during training \cite{2017WGAN_Arjovsky}. To address these issues, a variant called Wasserstein GAN with Gradient Penalty (WGAN-GP) is applied in this paper \cite{2017WGANGP_Ishaan}. WGAN-GP uses a different loss function based on the Wasserstein distance, leading to more stable training and better convergence properties. The loss function of WGAN-GP is shown in Equation~\ref{methodeq:WGANgp}, where $\mathbb{E}_{y \sim p(y)}[D_\omega(y)]$ represents average critic score correspond to real samples, which is supposed to be minimized by discriminator. $\mathbb{E}_{z \sim p_z(z)}[D_\omega(G_\theta(z))]$ represents average critic score assigned to generated samples $\hat{y}$ by discriminator, which should be maximized by discriminator and minimized by generator.

\begin{equation}
\begin{split}
    \min_{\theta}\max_{\omega}\mathcal{L}(D_\omega, G_\theta) = \mathbb{E}_{z \sim p_z(z)}[D_\omega(G_\theta(z))] - \mathbb{E}_{y \sim p(y)}[D_\omega(y)] \\
    + \lambda\mathbb{E}_{\hat{y} \sim p_{\hat{y}}}\left[(\lVert\nabla_{\hat{y}} D_\omega(\hat{y})\rVert_2 - 1)^2\right]
\end{split}\label{methodeq:WGANgp}
\end{equation}\par

The Wasserstein distance term $\mathbb{E}_{z \sim p_z(z)}[D_\omega(G_\theta(z))] - \mathbb{E}_{y \sim p(y)}[D_\omega(y)]$ measures the average difference between the critic scores assigned to generated samples and real samples. Specifically, it evaluates how far the distribution of generated samples is from the distribution of real samples by comparing their critic scores. Here, $\mathbb{E}_{y \sim p(y)}[D_\omega(y)]$ is the average score given by the critic to real samples, representing how "real" the critic perceives actual data, while $\mathbb{E}_{z \sim p_z(z)}[D_\omega(G_\theta(z))]$ is the average score assigned to generated samples, reflecting the critic’s assessment of the generated data's quality. The difference between these averages serves as an approximation of the Wasserstein distance. Unlike traditional GAN losses that rely on binary classification, the Wasserstein distance provides smooth and continuous gradients that guide the generator towards producing more realistic samples, making training more stable and less prone to issues like mode collapse.\par

The term $\lVert\nabla_{\hat{y}} D_\omega(\hat{y})\rVert_2$ represents the $L_2$ norm of the gradients of the discriminator with respect to interpolated samples $\hat{y}$. These interpolated samples are linearly mixed between real and generated data, ensuring that the gradient penalty applies along the path connecting these distributions. The significance of the $L_2$ norm being close to 1 lies in enforcing the Lipschitz continuity constraint—a mathematical condition necessary for the Wasserstein distance to be a valid approximation. This helps prevent unstable training dynamics by controlling the range of the discriminator’s gradients, avoiding issues where gradients become too large (exploding) or too small (vanishing). The hyperparameter $\lambda$ controls the weight of this gradient penalty term, balancing how strictly the model enforces this constraint. A higher $\lambda$ emphasizes maintaining the norm close to 1, which is critical for stability and ensures the critic does not overly exaggerate differences between real and generated data.\par

In this paper, we utilize a WGAN-GP variant combined with a DCGAN architecture specifically tailored for reconstructing 3D micro-CT images. This combination leverages the stable training properties of WGAN-GP and the convolutional strengths of DCGAN to effectively handle complex, grid-based data like volumetric images. Detailed descriptions of the generator and discriminator architectures, including parameter settings and layer configurations, can be found in the appendix.\par

We set the gradient penalty term, $\lambda$, to 10, consistent with the original WGAN-GP paper by Gulrajani et al.\cite{2017WGANGP_Ishaan}. For optimization, we used the Adam (Adaptive Moment Estimation) optimizer\cite{2014Kingma_Adam}. The discriminator's learning rate was set to $1e-4$, while the generator's learning rate was set to $5e-4$. A larger learning rate was assigned to the generator to expedite convergence and encourage the exploration of the generation distribution. In each training iteration, the discriminator was trained five times for every single training step of the generator. Training the discriminator more frequently allows it to better approximate the Wasserstein distance, providing more accurate gradients that guide the generator’s updates. This increased frequency helps the discriminator stay ahead of the generator, preventing the generator from exploiting weaknesses in the discriminator that could arise if both networks were trained equally often. This approach, recommended by Arjovsky et al.\cite{2017WGAN_Arjovsky} and Gulrajani et al.\cite{2017WGANGP_Ishaan}, enhances the stability of the training process, reduces the likelihood of mode collapse, and ensures the generator receives meaningful feedback, leading to improved generation quality.

\subsection{Gradual Perturbation}
\label{subsec:method/gradual_perturb}
As mentioned in section~\ref{subsec:method/GAN}, the Gaussian-based latent vector $z$ is upsampled by the generator to produce a synthetic image $\hat{y}$. Thus, many GAN-based conditional generation research efforts focus on building a secondary model, such as a neural network, to create a mapping between the latent space and image attributes \cite{2019chan_ganparameteric}. Another option is to perform a search process using a Markov chain or reinforcement learning (RL), to find the latent vector or stochastic injection parameter that imparts the correct conditioning characteristics to the resultant image. However, conditional generation can be accomplished more easily through a simpler approach that utilizes the characteristics of Gaussian random variables. Any mixture of Gaussian variables itself results in a Gaussian random variable .The traditional Gaussian-based geostatistical model calibration technique - the gradual Gaussian deformation approach, cleverly utilizes this property of Gaussian variables. Embedding this approach within the WGAN process results in a conditioning approach that is more flexible compared to RL but not as computationally expensive as Markov chain Monte Carlo (MCMC) methods. Within an inversion workflow, Hu (2000) \cite{2000Liu_graudalperturb} proposed gradual deformation and iterative calibration of Gaussian-related stochastic geology models to yield reservoir models that match with final production response. A  perturbation combines two independent Gaussian random functions~($z_1$ and $z_2$) with an identical covariance function, as described in equation~\ref{methodeq:combineGaussian}. The combined random function $z(t)$ can be considered a new Gaussian realization, with a tuning parameter $t$ that can be calibrated by defining an objective function $O = f(z(t))-R$, where $R$ is our real observation response and $f(z(t))$ is the response obtained by applying a forward physical model.\par
\begin{equation}
    z(t) = z_1 cos(t) + z_2 sin(t)
    \label{methodeq:combineGaussian}
\end{equation}

GAN training is completed in the first stage. The second stage involves using the gradual Gaussian deformation to perturb the Gaussian latent vector to generate 3D synthetic microstructures that match the target physical properties. The general workflow of such controllable generation is depicted in figure~\ref{method_fig:GAN_workflow}. The target physical properties can be derived from well logs or field scale geologic characterization models so that conditional generation of 3D microstructures is consistent with the variations in larger-scale geological properties. The physical properties of the micro-scale model can be easily calculated using a forward physical simulator such as a pore network model. In the context of GAN-based generation, $z(t)$ becomes the Gaussian latent vector for the generator during the optimization process, and our forward physical model $f(z(t))$ is a pore network model that can simulate multiple rock properties from reconstructed 3D porous media, such as porosity, permeability, mean pore size distribution, etc. The calculated properties are compared against those derived from well logs.

Minimizing the objective function using a single epoch is insufficient; instead, an iterative approach must be implemented to obtain a continuous chain of realizations $z_n(t)$, as described in equation~\ref{methodeq:iterative_gaussian}, where $t$ is the perturbation parameter. The iteration will continue until the target physical property is achieved. In reality, it's not possible to generate porous media that exactly match the user input property, thus a certain error threshold needs to be defined for different physical attributes. The iterations are continued until this threshold is reached. To ensure a more efficient iteration process, stochastic gradient descent is used to calibrate perturbation. The simulation mismatch error $e=R-\hat{R}$ will be directly backpropagated to the Gaussian vector under the assumption that vector is fully responsible for the sensitivity of physical simulation results. The step-by-step calibration process in our case is comprehensively delineated in Algorithm~\ref{methodalg:gaussianperturbation}.

\begin{equation}
    z_n(t) = z_{n-1} cos(t) + u_n sin(t)
    \label{methodeq:iterative_gaussian}
\end{equation}

\begin{algorithm}
\caption{Gradual Gaussian Perturbation Guided by Physical Forward Model}
\label{methodalg:gaussianperturbation}
\begin{algorithmic}[1]
\ENSURE{$G_\theta$ is a pretrained model of WGAN-GP}
\STATE{$f(G_\theta)$: pore network model on generated images by $G_\theta$}
\STATE{Initialize Gaussian vector $z_{n-1}\sim N(0,I)$}
\STATE{Define target physical property as $R$}
\STATE{Define final error acceptance threshold as threshold $\gamma$}
\STATE{Initialize error to be $Inf$}
\STATE{Initialize learning rate to be $\eta$}
\STATE{Define objective function $O = \frac{1}{2} (R-\hat{R})^2$}

\WHILE{$e>\gamma$}
    \STATE{Sample parameter t from $U(0,2\pi)$} 
    \STATE{In step $n$, sample random Gaussian vector $u_n \sim N(0,I)$}
    \STATE{In step $n$, Obtain perturbed Gaussian vector $z_n(t)=z_{n-1}cos(t)+u_n sin(t)$}
    \STATE{Generate image $\hat{y} = G_\theta(z_n(t))$ based on perturbed Gaussian vector $z_n(t)$}
    \STATE{Forward model on generated image to obtain estimated property $\hat{R}=f(\hat{y})$}
    \STATE{Calculate error as $e = abs(R-\hat{R})$}
    \STATE{Gradient calculation $\frac{\partial O}{\partial f} \frac{\partial z_n}{\partial t}$} and update $t$ using gradient descent $t = t-\eta \frac{\partial O}{\partial f} \frac{\partial z_n}{\partial t}$
    \STATE{Update $z_n(t)$}
    \STATE{Replace $z_{n-1}$ with updated $z_n$}
\ENDWHILE
\end{algorithmic}
\end{algorithm}\par


We used OpenPNM\cite{2016openpnm}, a Python-based pore network model simulation package, to build the physical simulator. However, this framework can be used with any physical simulation approach such as the Lattice Boltzmann method as long as it is computationally efficient enough to go through several iterations (you can check the number of iterations for different physical properties in section~\ref{sec:results}). This optimization framework does not require a fully differentiable physical model concatenated with our pretrained generator to gradually calibrate generative realizations since the Gaussian realizations of latent vectors in GAN can be directly calibrated by physical simulation responses. Meanwhile, any physical attributes simulated by the pore network model can be used to generate corresponding matched porous media. These rock properties-constrained porous media can serve as representative porous media which share similar attributes compared to larger-scale models, such as well or field scale models, and multiphase flow properties can potentially be upscaled through this process.

\begin{figure}[hbt]
    \centering
    \includegraphics[width = \textwidth]{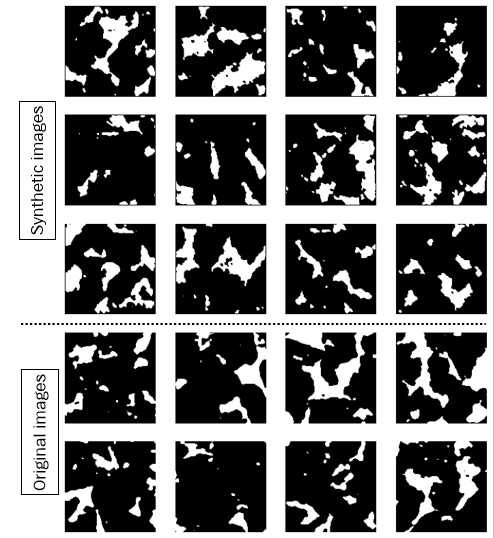}
    \caption{2D Slices of Synthetic Images vs. 2D Slices of Original Training Images
}
    \label{result_fig:real_img_vs_fake_img}
\end{figure}\par

\subsection{Evaluation Criteria}
There are mainly two ways to assess the performance of our modeling framework. Firstly, we evaluate the quality of the reconstruction in terms of the connectivity of the resultant models. Secondly, we assess how accurately and efficiently the Gradual Gaussian Perturbation can control the GAN generation process given user-defined rock properties.\par

To evaluate reconstruction quality, besides visual inspection, we have devised a set of metrics that compare the physical statistics of training images with those of synthetic images. These metrics are similar to the evaluation metrics used by Mosser, 2017\cite{2017MosserGAN}. The physical properties associated with the pore network pertain to the intrinsic characteristics of porous media, that can be described using Minkowski functionals that can be used to characterize the topology of the pore structure, such as porosity ($\phi$), average curvature, and Euler characteristics ($\chi$)\cite{2000Mecke:Minkowski}.

\alphsection{Porosity $\phi$}
The first Minkowski functional of order zero is porosity, which is defined as the ratio of the void space volume to the bulk volume of porous media, as described in Equation~\ref{methodeq:phi}. This property is also very commonly observed at both well and field scales and can be considered as one of the key attributes to link pore-scale models and field-scale observations. We will use this variable to evaluate both GAN generation quality and the ability to condition the models.
\begin{equation}
\phi = \frac{V_p}{V}
\label{methodeq:phi}
\end{equation}

\alphsection{Specific area}
The specific surface area, which represents the first-order Minkowski functional, can be calculated using Equation~\ref{methodeq:spec_area}, where $S_v$ denotes the void-solid interface. This parameter plays a critical role in characterizing the morphological properties of porous media and is related to absolute permeability through the Carman-Kozeny equation.

\begin{equation}
A_p = \frac{S}{V}
\label{methodeq:spec_area}
\end{equation}

\begin{equation}
    S_v = \frac{1}{V}\int dS
\end{equation}\label{methodeq:spec_area}

\alphsection{Euler Characteristic $\chi_p$}
The Euler characteristic represents the third-order Minkowski functional and can be employed to quantify static pore-to-pore connectivity. It can be calculated from micro-CT images or reconstructed models using Equation~\ref{methodeq:eul}, where $N$ denotes the number of isolated objects, $L$ signifies redundant loops, and $O$ represents the number of cavities.

\begin{equation}\label{methodeq:eul}
\chi_p = N-L+O
\end{equation}

\alphsection{Absolute permeability $k_{abs}$}
The absolute permeability ($k_{abs}$) is determined using Darcy's Law, as outlined in the system of equations given by~\ref{methodeq:kabs} through stokes flow simulation. It is a crucial parameter to quantify transport properties in a porous medium. Like porosity, it is also a crucial parameter to link pore-scale models to field-scale characterization.

\begin{equation}
\label{methodeq:kabs}
\begin{aligned}
\nabla \cdot \boldsymbol{u} &= 0 \\
-\nabla p + \mu \nabla^2 \boldsymbol{u} &= \boldsymbol{0} \\
\frac{\mu L Q}{A \Delta P} &= k_{abs}
\end{aligned}
\end{equation}

where $\boldsymbol{u}$ is the fluid velocity, $p$ is the pressure, $\mu$ is the fluid viscosity, $L$ is the length of the porous medium, $Q$ is the volumetric flow rate, $A$ is the cross-sectional area, and $\Delta P$ is the pressure drop.
The first equation in the system represents the continuity equation for incompressible flow, stating that the divergence of the velocity field is zero. The second equation is the Stokes equation, which describes the balance between pressure gradient and viscous forces in low Reynolds number flow. The third equation is the expression for Darcy's Law  that expresses absolute permeability as a parameter relating the flow rate to  pressure drop, and fluid properties.

\alphsection{Mean pore size $\Bar{D_p}$ and mean throat size $\hat{D_t}$}

Mean pore size and mean throat size are crucial pore structure parameters that influence relative permeability. They will be used as important reconstruction evaluation metric as well as for conditional model generation.


\section{Results}
\label{sec:results}

\subsection{Reconstruction Quality}
The dataset used for training and evaluating the GAN is pore scan data for Berea sandstone\cite{2020IBMresearch}, which is an open-source dataset provided by IBM research. The original size of the scan image is $1000^3$ voxels with a resolution of $2.25 \mu m$. However, training a GAN model on a scale of $1000^3$ voxel image is not efficient in terms of computation and availability of training samples. Thus, the original 3D cube is cropped into smaller 3D subvolumes on a scale of $128^3$ voxels. At that scale, porosity variations between approximately $0.15 \sim 0.3$ can be observed across the sub-volumes. A scale larger than $128$ would result in less variability in the porosity of samples, increase the computational cost of training the GAN, and effectively decrease the number of training samples. To ensure that the number of training samples is large enough, they are sampled as overlapping volumes with a shift of approximately $30$ voxels to generate a total of 23,839 $128^3$ 3D volumes. The training history and hardware information can be found in the appendix.\par

\begin{figure}[hbt]
  \centering
  \begin{minipage}{0.3\textwidth}
    \includegraphics[width=\linewidth]{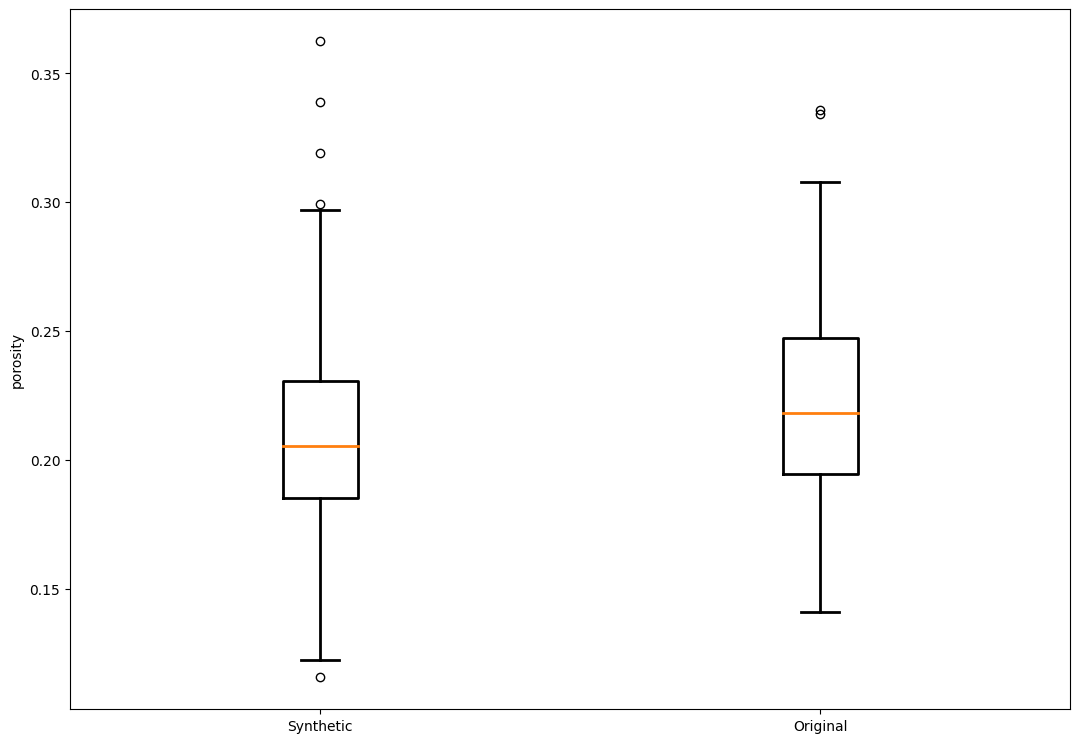}
    \caption*{(a) $\phi$}
  \end{minipage}\hfill
  \begin{minipage}{0.3\textwidth}
    \includegraphics[width=\linewidth]{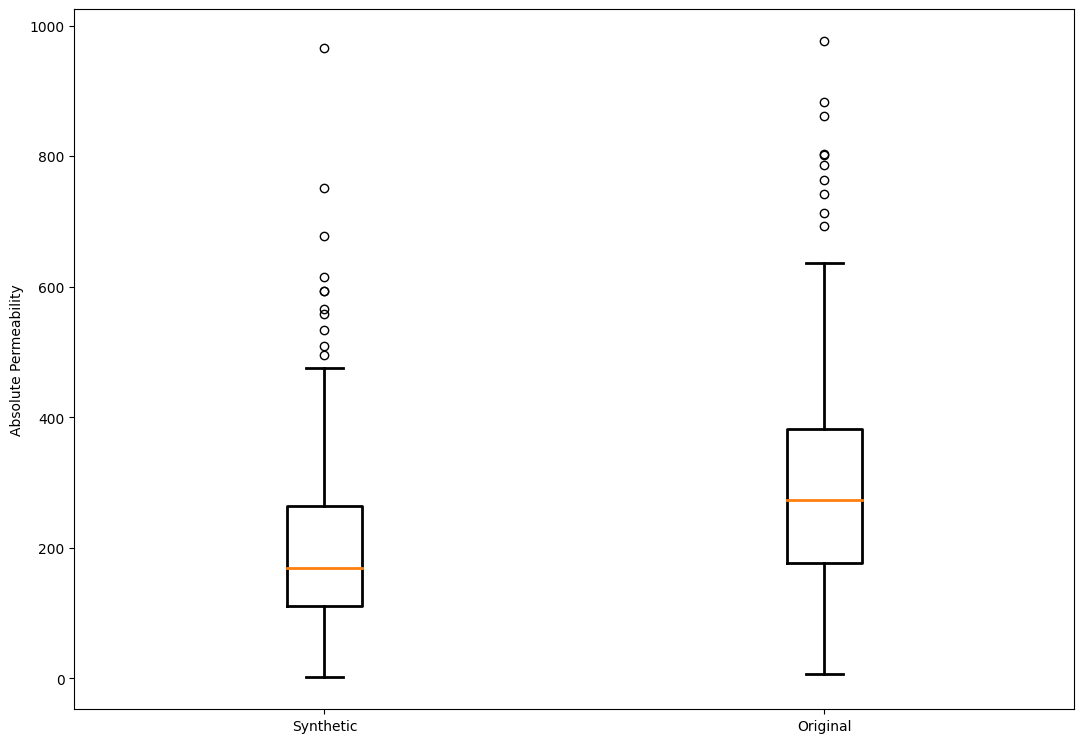}
    \caption*{(b) $k_{abs}$}
  \end{minipage}\hfill
  \begin{minipage}{0.3\textwidth}
    \includegraphics[width=\linewidth]{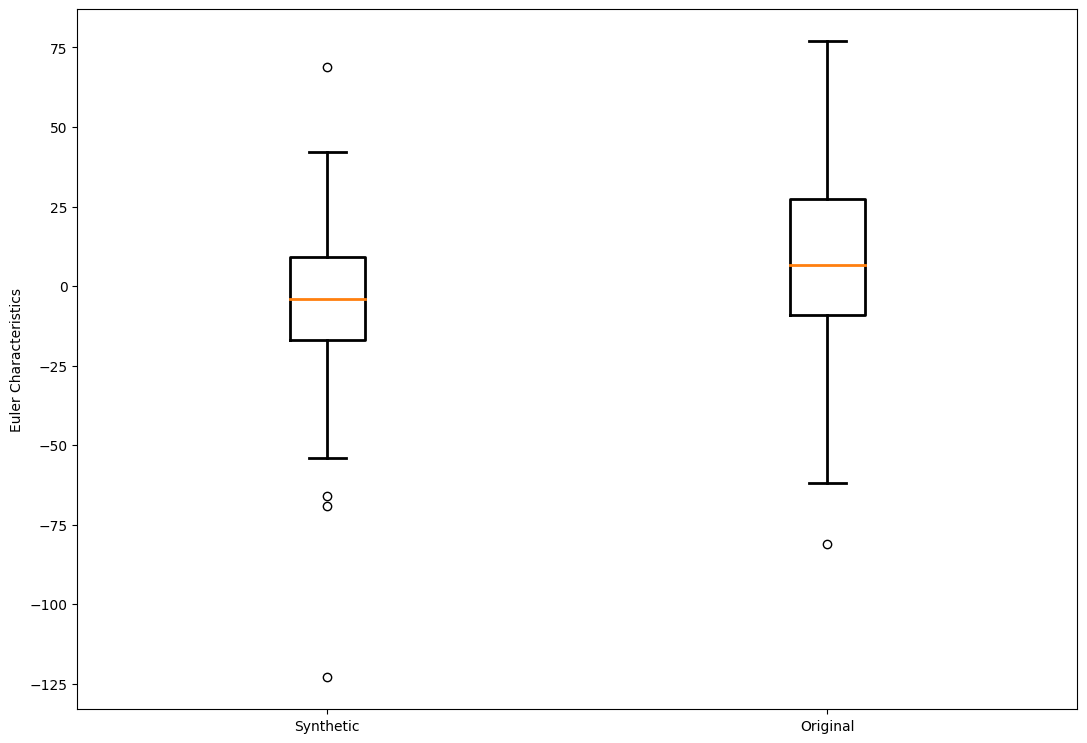}
    \caption*{(c) $\chi_p$}
  \end{minipage}
  \vspace{1cm} 
  \begin{minipage}{0.3\textwidth}
    \includegraphics[width=\linewidth]{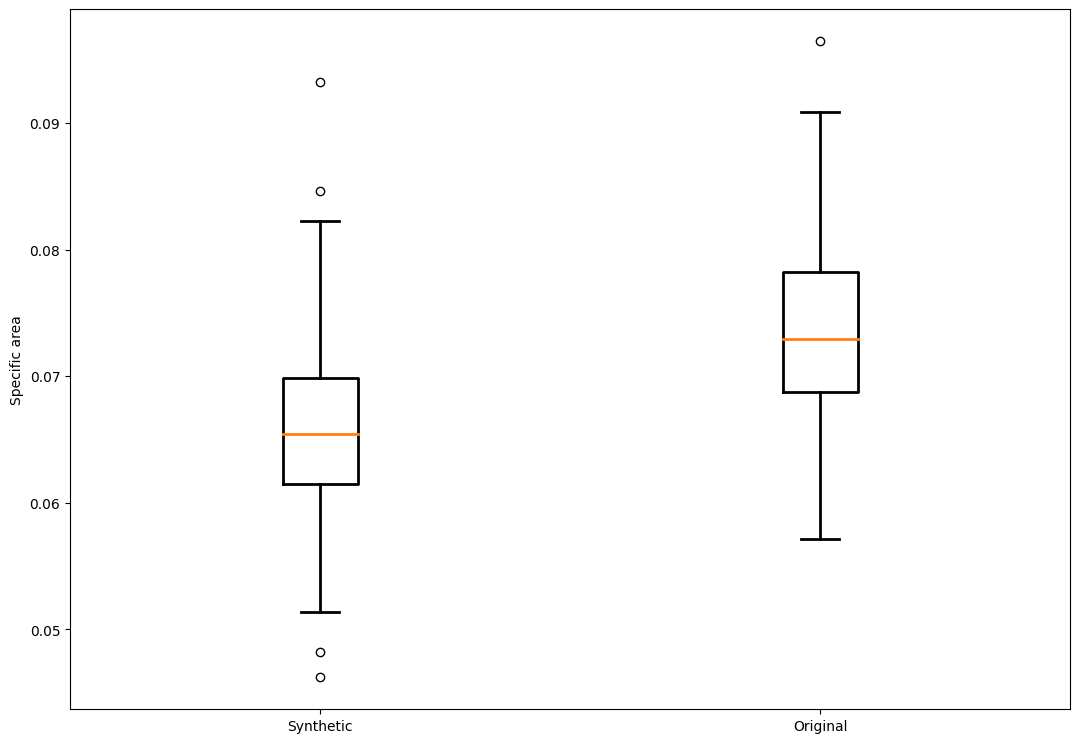}
    \caption*{(d) Specific area}
  \end{minipage}\hfill
  \begin{minipage}{0.3\textwidth}
    \includegraphics[width=\linewidth]{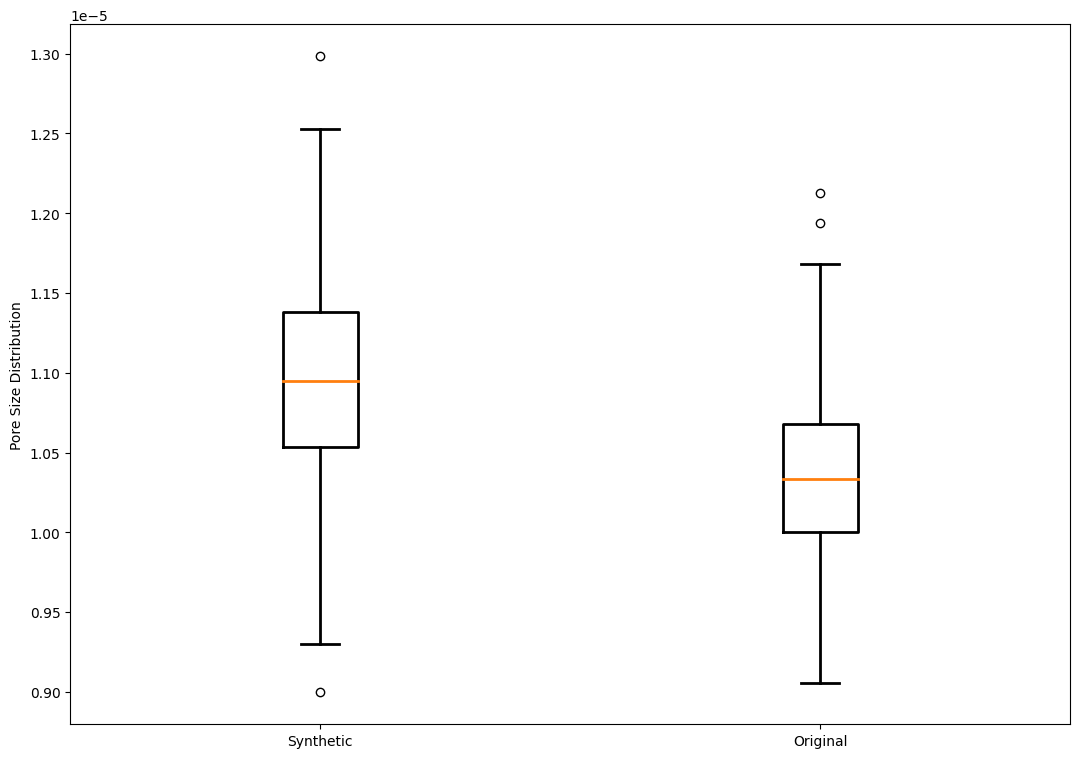}
    \caption*{(f) $\Bar{D_p}$}
  \end{minipage}\hfill
  \begin{minipage}{0.3\textwidth}
    \includegraphics[width=\linewidth]{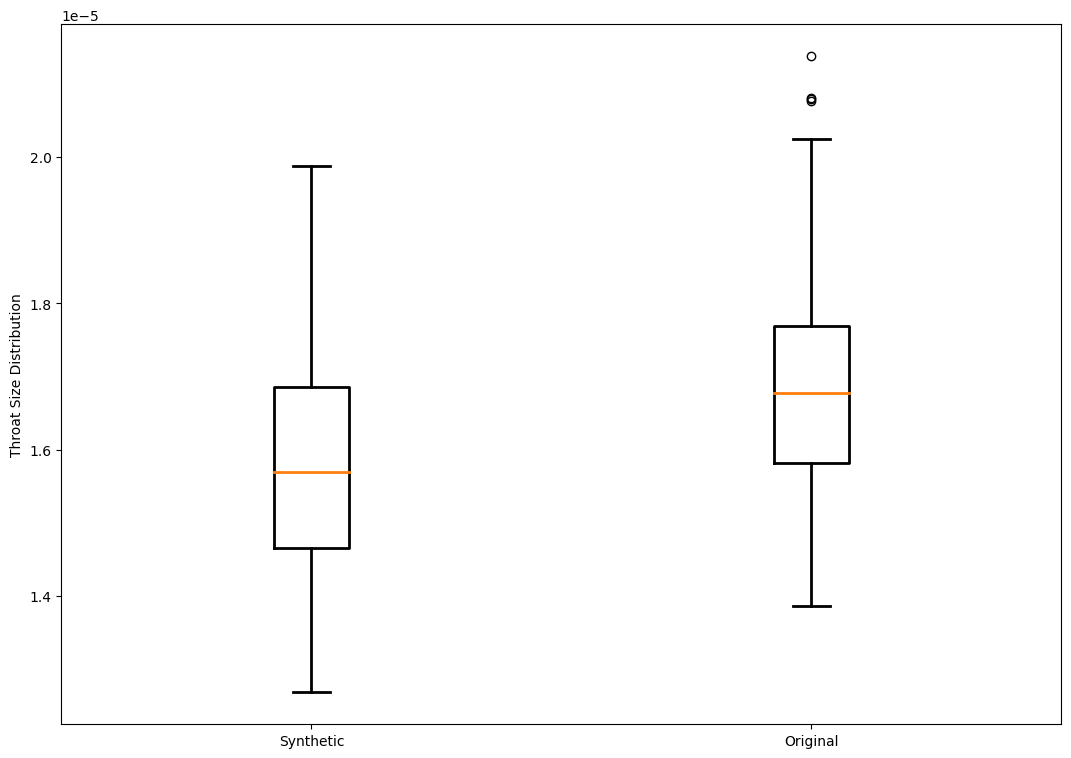}
    \caption*{(g) $\Bar{D_t}$}
  \end{minipage}
  \caption{Box plots of statistics showing the spread in properties over the original training set (left) as well as the constructed synthetic samples (right).}
  \label{result_fig:6boxplots}
\end{figure}

The following section will focus on evaluating the quality of the GAN-reconstructed 3D volumes. We first assess the reconstruction quality by visually comparing synthetic images with original training images, as demonstrated in Figure~\ref{result_fig:real_img_vs_fake_img}. GAN-generated 3D microstructure images are post-processed through a median filter and Multi-Otsu Thresholding image processing technique to convert the generated image tensor to a boolean matrix for pore network modeling\cite{2001Liao_Ostu}. The synthetic images appear remarkably similar to the training images, although with some tiny fragments observed in the synthetic generated samples. Overall, the synthetic images successfully capture the voxel patterns of the Berea sandstone microstructure.\par

To further quantitatively evaluate the reconstruction quality, we compare the physical properties of the training images with those of the synthetic images (300 samples). As described in Section \ref{sec:Method}, the physical properties considered include porosity ($\phi$), specific surface area ($A_p$), Euler characteristic ($\chi_p$), absolute permeability ($k_\mathrm{abs}$), mean pore size ($\bar{D}p$), and mean throat size ($\bar{D}_t$). These properties are calculated for both the training and synthetic images, and their distributions are compared using box plots, as shown in Figure \ref{result_fig:6boxplots}. We observe that the properties computed on synthetic images exhibit slightly broader variability compared to those computed using the training samples. In terms of mean, quartile, and extreme values, the synthetic statistics and original statistics generally fall within the same range for most physical properties. However, the synthetic $k_\mathrm{abs}$ distribution appears to be consistently lower than the original $k_\mathrm{abs}$ distribution, as does the specific surface area. This discrepancy may be attributed to the GAN-based algorithm's limitations in reconstructing the curvature and shape of pore surfaces, which can impact derived properties such as $k_\mathrm{abs}$ and specific surface area. Reconstructing the correct curvature is more challenging than accurately reproducing pore volume, as reflected in properties like porosity. This discrepancy also points to a future research direction that focuses on the coherency of voxel patterns when reconstructing the porous media.

\begin{figure}[hbt!]
\centering
    \begin{minipage}[t]{0.4\textwidth}
    \includegraphics[width=\textwidth]{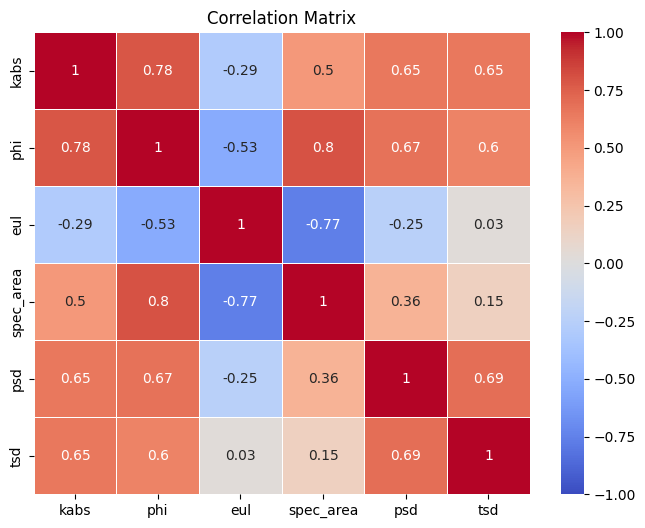}
    \caption{ Correlation map between pore properties calculated on the  trining set.}
    \label{result_fig:corr_real}
    \end{minipage}
    \hspace{0.05\textwidth}
    \begin{minipage}[t]{0.4\textwidth}
    \includegraphics[width=\textwidth]{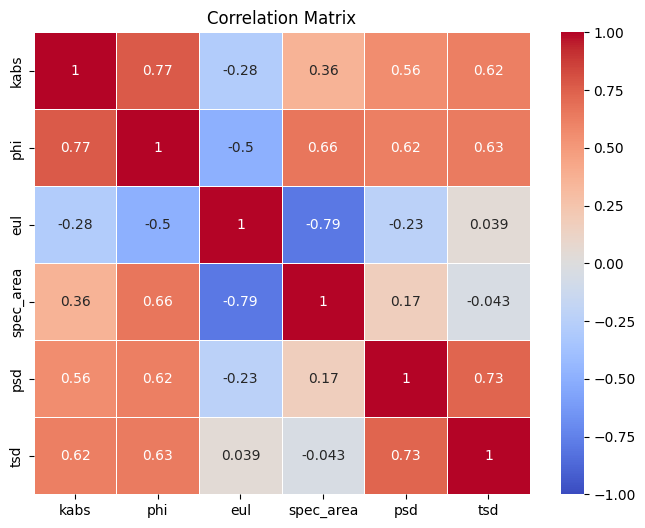}
    \caption{Correlation map between pore properties calculated on the  generated images.}
    \label{result_fig:corr_fake}
    \end{minipage}
\end{figure}\par

\begin{figure}[hbt]
    \centering
    \includegraphics[width = 0.8\textwidth]{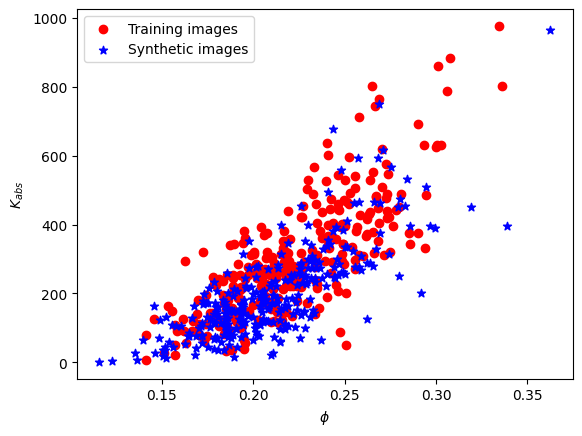}
    \caption{Porosity vs permeability~($md$)}
    \label{result_fig:kvsphi}
\end{figure}\par

It is essential not only to examine the quality of the reconstructed pore models in terms of the closeness of the computed properties to the original statistics but also to investigate whether correlations among these physical properties have been preserved. We analyze and plot the correlation matrix for both the training samples and synthetic samples by calculating the Pearson correlation coefficient, as described in Equation~\ref{result_eq:corr}. In this equation, $x_i$ and $y_i$ represent the individual values for properties $x$ and $y$, while $\bar{x}$ and $\bar{y}$ denote the means of properties $x$ and $y$, respectively. The corresponding correlation matrix heatmaps for the original and synthetic physical properties can be found in Figure~\ref{result_fig:corr_real} and Figure~\ref{result_fig:corr_fake}.

\begin{equation}
  r = \frac{\sum_{i} (x_i - \bar{x})(y_i - \bar{y})}{\sqrt{\sum_{i}(x_i - \bar{x})^2} \cdot \sqrt{\sum_{i}(y_i - \bar{y})^2}}
    \label{result_eq:corr}
\end{equation}\par

When comparing the property correlation matrices for the synthetic and original pore networks, most properties exhibit similar correlation trends. It is encouraging to observe that the relationship between porosity and absolute permeability has been well preserved: originally at 0.78 and at 0.77 for the synthetic models. We also create a scatter plot to compare the porosity and permeability relationships both among the training samples and the synthetic samples, as shown in Figure~\ref{result_fig:kvsphi}. The $k$-$\phi$ trend is mostly well preserved, except for a slightly smaller spread of values observed in the original training set.\par

\subsection{Conditional results}

Based on previous results and evaluations of the pretrained Wasserstein DCGAN-GP, most physical attributes and the correlations between them have been well preserved leading us to conclude that the quality of reconstruction is good. In this section, we evaluate the accuracy and efficiency of implementing the Gradual Gaussian Deformation approach to condition the GAN to user specified rock properties as described in Section~\ref{sec:Method}. Our generated images are constrained based on four physical attributes: porosity $\phi$, absolute permeability $k_\text{abs}$, mean pore size parameter $D_p$, and mean throat size parameter $D_t$. We reconstruct approximately a hundred 3D microstructure models conditioned on these properties within certain ranges and optimization thresholds. The optimization thresholds (which serve as stopping criteria for gradual perturbation of the GAN's latent vector) and target ranges for various geological properties are defined as follows:

\begin{itemize}
    \item Porosity: $\phi \pm 0.01$ within a uniform distribution range $\mathcal{U}(0.14, 0.30)$
    \item Absolute permeability: $k_\text{abs} \pm 15\,\text{mD}$ within a uniform distribution range $\mathcal{U}(100, 300)$
    \item Mean pore size diameter: $\bar{D_p} \pm 1 \times 10^{-7}\,\text{m}$ within a uniform distribution range $\mathcal{U}(0.95 \times 10^{-5}, 1.1 \times 10^{-5})$
    \item Mean throat size parameter: $\bar{D_t} \pm 5 \times 10^{-8}\,\text{m}$ within a uniform distribution range $\mathcal{U}(3.6 \times 10^{-6}, 4.2 \times 10^{-6})$
\end{itemize}

\begin{figure}[hbt]
  \centering
  \begin{minipage}{0.45\textwidth}
    \centering
    \includegraphics[width=\linewidth]{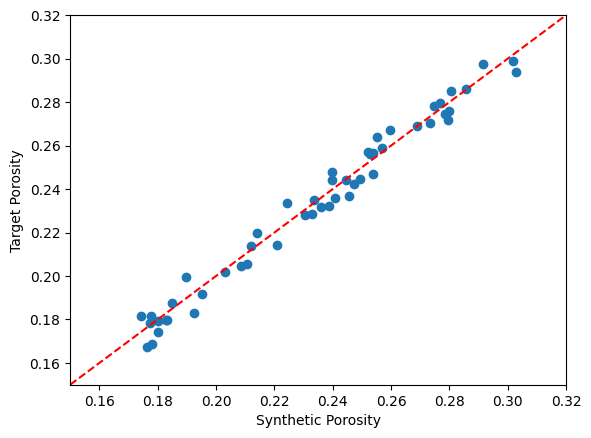}
    \caption{Scatter plot of synthetic porosity versus target porosity with normalized $\text{RMSE} = 0.04$. Each realization takes approximately 13.2 seconds to generate a 3D microstructure that preserves the target porosity.}
    \label{result_fig:phivsphi}
  \end{minipage}\hfill
  \begin{minipage}{0.45\textwidth}
    \centering
    \includegraphics[width=\linewidth]{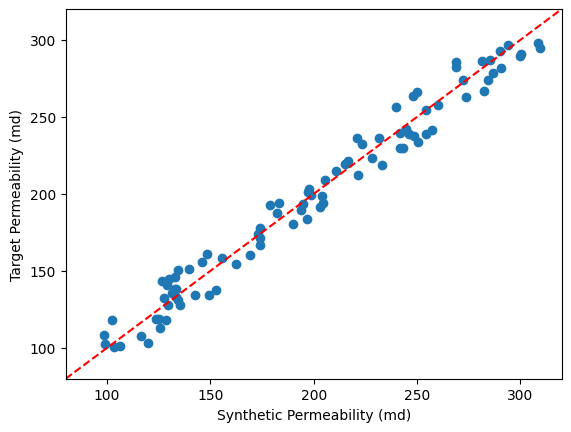}
    \caption{Scatter plot of synthetic permeability versus target permeability with normalized $\text{RMSE} = 0.05$. Each realization takes approximately 42 seconds to generate a 3D microstructure that preserves the target permeability.}
    \label{result_fig:kvsk}
  \end{minipage}
  \medskip
  \begin{minipage}{0.45\textwidth}
    \centering
    \includegraphics[width=\linewidth]{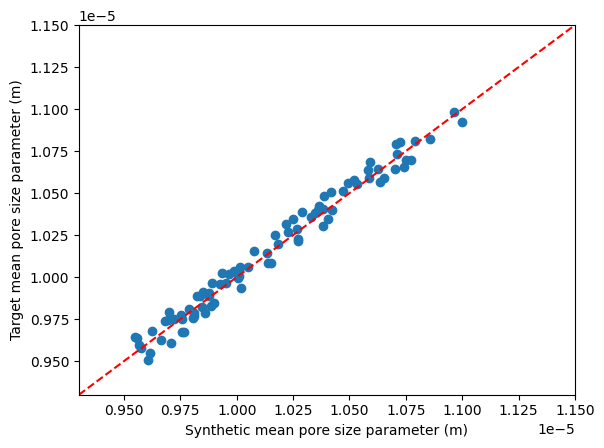}
    \caption{Scatter plot of synthetic mean pore size versus target mean pore size with normalized $\text{RMSE} = 0.04$. Each realization takes approximately 102 seconds to generate a 3D microstructure that preserves the target mean pore size.}
    \label{result_fig:psdvspsd}
  \end{minipage}\hfill
  \begin{minipage}{0.45\textwidth}
    \centering
    \includegraphics[width=\linewidth]{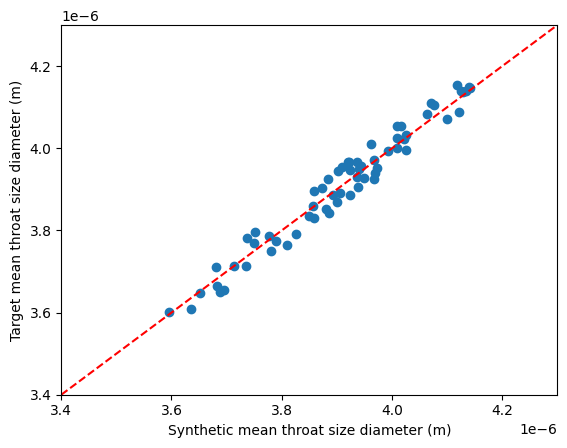}
    \caption{Scatter plot of synthetic mean throat size versus target mean throat size with normalized $\text{RMSE} = 0.05$. Each realization takes approximately 5 minutues to generate a 3D microstructure that preserves the target mean throat size.}
    \label{result_fig:tsdvstsd}
  \end{minipage}
\end{figure}

We generated synthetic porous media using gradual Gaussian perturbation based on the above conditioning criteria, and the comparison between the simulated properties and the target conditioning values are shown in Figures~\ref{result_fig:phivsphi} to~\ref{result_fig:tsdvstsd}. We used \em{OpenPNM} as our pore network modeling simulation software \cite{2016openpnm} and \em{porespy} as our image processing tool \cite{2019porespy}. The Gaussian deformation approach can accurately reconstruct porous media conditioned to target property ranges with quite low RMSE, which has been normalized by the range of the target property. In terms of efficiency, porosity-driven microstructure generation has the lowest time per sample to generate target porous media. This is because porosity calculation does not require pore network modeling simulation. The generation driven by other physical properties all require pore network modeling, which is more computationally expensive. However, overall the method remains very efficient, as even the permeability-constrained microstructure generation only takes 42 seconds per optimization. For all four properties, this framework requires an average of approximately 15 epochs of pore network modeling or image processing to find the target properties. These results indicate that by employing a physics-informed gradual Gaussian perturbation approach, we can successfully perturb porous media while conditioning them to specific user-defined  physical properties.

\section{Conclusion}
\label{sec:conclusion}

We have successfully implemented an unsupervised learning GAN-based 3D porous media reconstruction workflow that learns information from segmented micro-CT image cubes. The GAN model is trained using a pair of Deep Convolutional Generative Adversarial Networks. The Wasserstein distance is used as the discriminator loss function while forcing the gradient of the discriminator's output with respect to its input to have a norm close to 1. We developed several physical evaluation metrics to compare the reconstruction quality. The results show that the GAN-based reconstructions agree well with the training images in terms of the statistics of rock and pore-scale properties computed on the training and synthetic images. The correlations between properties are also well reproduced in the synthetic images. Although the GAN may not perfectly learn the curvature of pore shapes from training images, the results show that the reconstructed images are of good overall quality.\par

For controllable generation, we introduced a Physics-Informed Gradual Gaussian Deformation approach that gradually perturbs the latent space within Wasserstein DCGAN. The reproduction of the target property is evaluated using a pore network model embedded within an optimization loop. The established conditioning scheme can control GAN generation constrained by any physical property extracted from the pore network model. In our case, we use porosity, absolute permeability, mean pore size, and mean throat size of pores as our physical constraints. The results, as illustrated in Figures~\ref{result_fig:phivsphi} to~\ref{result_fig:tsdvstsd}, demonstrate the effectiveness of the method for reconstructing 3D porous media. There are several advantages to using the physics-informed iterative Gaussian perturbation approach to control GAN generation of the 3D micro-structures in rocks:

\begin{itemize}
    \item The proposed conditioning scheme can be used to condition the pore-scale models to any physical property that can be extracted from a forward model, such as a pore network model. The whole process requires no post-training efforts. This framework does not modify the original base model based on Wasserstein DCGAN-GP. Once the generator is trained successfully, the process of conditioning can be achieved using any forward modeling process, such as direct image simulation using the Lattice Boltzmann method. 
    \item The proposed conditioning approach does not require a differentiable physical simulator, although having one could increase the optimization efficiency within the gradual deformation scheme.
\end{itemize}

\section{Limitation}

Despite the success of our GAN-based approach for porous media reconstruction, there are several limitations that need to be addressed:

\begin{itemize}
    \item \textbf{Fixed reconstruction size:} The current implementation generates synthetic samples at a fixed size, which is determined by the architecture of the GAN. Although current scale~($128^3$) may be at the representative elementary volume~(REV) scale for porosity, it may be too small for reliable prediction of more complex flow functions such as permeability and relative permeability that tend to have larger REV scales \cite{2020Jackson_krREV}. The computational cost constrains our ability to produce larger-scale reconstructions or to adapt the output size based on different representative scales physical properties. Future work could explore techniques such as auto-regressive based multi-scale architectures to overcome this limitation.
    
    \item \textbf{Lack of heterogeneity control:} While our approach successfully reproduces various physical properties, it is not easy to extend the method to reflect spatial heterogeneity within the porous media. The generated samples maintain statistical similarity to the training data but may not capture larger-scale variations or specific spatial patterns that may be observed in real rock formations.
    
    \end{itemize}

It is important to note that while our generator can produce physically reasonable synthetic samples that are constrained to user-defined or field inferred rock properties, it may not be adequate for completing more complex tasks such as performing scale up of rock properties. In order to perform such complex tasks, further development may be needed to condition the pore model generation process to available field-scale data incorporating larger-scale spatial variability that may be observed in subsurface geologic formations.

\section*{Acknowledgments}
This work was partially supported by the Carbon Storage (SMART-CS) Initiative funded by the U.S. Department of Energy’s (DOE) Office of Fossil Energy’s Carbon Storage Research program through the National Energy Technology Laboratory (NETL).

\bibliographystyle{unsrt}  
\bibliography{references}

\newpage

\section*{Appendix}

The training history of the WGAN-GP model with a DCGAN-based architecture is presented in Figure~\ref{appendix_fig:WGANGP_trainhist} in the appendix. The model was trained using an NVIDIA RTX A6000. As observed, the discriminator and generator losses converge effectively. With an increase in epochs, the reconstructed 3D Micro-CT images become increasingly similar to the training images, to the point where they are nearly indistinguishable. The generator architecture has been detailed in Table~\ref{appendixtab:generator_architecture}, and the discriminator architecture has been detailed in Table~\ref{appendixtab:discriminator_architecture}.\par

\begin{figure}[hbt]
    \centering
    \includegraphics[width = \textwidth]{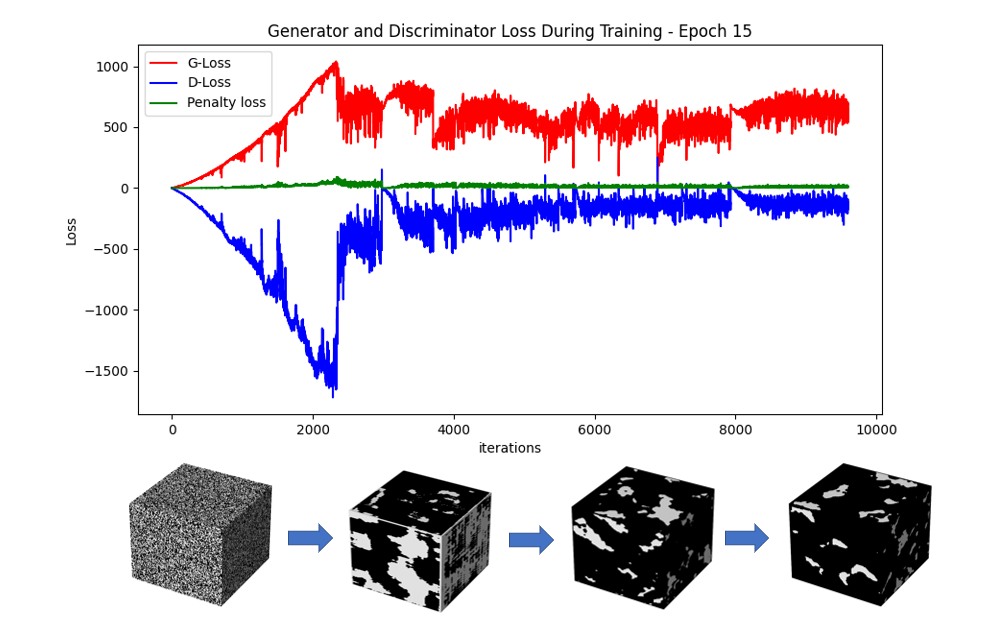}
    \caption{Training history of WGAN-GP combined with DCGAN}
    \label{appendix_fig:WGANGP_trainhist}
\end{figure}\par

\begin{table}[hbt]
\centering
\begin{tabular}{@{}lll@{}}
\toprule
Layer (type:depth-idx) & Output Shape & Param \# \\ \midrule
Generator & [5, 1, 128, 128, 128] & -- \\
$\vert-$Linear: 1-1 & [5, 131072] & 2,752,512 \\
$\vert-$BatchNorm1d: 1-2 & [5, 131072] & 262,144 \\
$\vert-$LeakyReLU: 1-3 & [5, 131072] & -- \\
$\vert-$Sequential: 1-4 & [5, 128, 16, 16, 16] & -- \\
$\vert\quad\vert-$ConvTranspose3d: 2-1 & [5, 128, 16, 16, 16] & 2,097,280 \\
$\vert\quad\vert-$BatchNorm3d: 2-2 & [5, 128, 16, 16, 16] & 256 \\
$\vert\quad\vert-$LeakyReLU: 2-3 & [5, 128, 16, 16, 16] & -- \\
$\vert-$Sequential: 1-5 & [5, 64, 32, 32, 32] & -- \\
$\vert\quad\vert-$ConvTranspose3d: 2-4 & [5, 64, 32, 32, 32] & 524,352 \\
$\vert\quad\vert-$BatchNorm3d: 2-5 & [5, 64, 32, 32, 32] & 128 \\
$\vert\quad\vert-$LeakyReLU: 2-6 & [5, 64, 32, 32, 32] & -- \\
$\vert-$Sequential: 1-6 & [5, 32, 64, 64, 64] & -- \\
$\vert\quad\vert-$ConvTranspose3d: 2-7 & [5, 32, 64, 64, 64] & 131,104 \\
$\vert\quad\vert-$BatchNorm3d: 2-8 & [5, 32, 64, 64, 64] & 64 \\
$\vert\quad\vert-$LeakyReLU: 2-9 & [5, 32, 64, 64, 64] & -- \\
$\vert-$Sequential: 1-7 & [5, 1, 128, 128, 128] & -- \\
$\vert\quad\vert-$ConvTranspose3d: 2-10 & [5, 1, 128, 128, 128] & 2,049 \\
$\vert\quad\vert-$Tanh: 2-11 & [5, 1, 128, 128, 128] & -- \\ \bottomrule
\end{tabular}
\caption{Generator Architecture. The generator architecture primarily consists of five neural network layers. (1) A linear layer that transforms the input noise vector into a high-dimensional representation, with a BatchNorm1d layer for normalization followed by a LeakyReLU activation function for non-linear activation. (2) A series of Sequential blocks, each containing a ConvTranspose3d (also known as deconvolution) layer, a BatchNorm3d layer for normalization, and a LeakyReLU activation function for non-linear activation. These blocks help in progressively upsampling the feature maps, while decreasing the number of channels (filters). The final Sequential block contains a ConvTranspose3d layer to produce the output feature map with the desired dimensions, followed by a Tanh activation function to constrain the output values within a specific range. In total, the generator has 5,769,889 parameters.}
\label{appendixtab:generator_architecture}
\end{table}

\begin{table}[hbt]
\centering
\begin{tabular}{@{}lll@{}}
\toprule
Layer (type:depth-idx) & Output Shape & Param \# \\ \midrule
Discriminator & [5, 1] & -- \\
$\vert-$ModuleList: 1-1 & -- & -- \\
$\vert\quad\vert-$Conv3d: 2-1 & [5, 4, 128, 128, 128] & 112 \\
$\vert\quad\vert-$InstanceNorm3d: 2-2 & [5, 4, 128, 128, 128] & 8 \\
$\vert\quad\vert-$LeakyReLU: 2-3 & [5, 4, 128, 128, 128] & -- \\
$\vert\quad\vert-$MaxPool3d: 2-4 & [5, 4, 64, 64, 64] & -- \\
$\vert\quad\vert-$Conv3d: 2-5 & [5, 16, 64, 64, 64] & 1,744 \\
$\vert\quad\vert-$InstanceNorm3d: 2-6 & [5, 16, 64, 64, 64] & 32 \\
$\vert\quad\vert-$LeakyReLU: 2-7 & [5, 16, 64, 64, 64] & -- \\
$\vert\quad\vert-$MaxPool3d: 2-8 & [5, 16, 32, 32, 32] & -- \\
$\vert\quad\vert-$Conv3d: 2-9 & [5, 64, 32, 32, 32] & 27,712 \\
$\vert\quad\vert-$InstanceNorm3d: 2-10 & [5, 64, 32, 32, 32] & 128 \\
$\vert\quad\vert-$LeakyReLU: 2-11 & [5, 64, 32, 32, 32] & -- \\
$\vert\quad\vert-$MaxPool3d: 2-12 & [5, 64, 16, 16, 16] & -- \\
$\vert-$Sequential: 1-2 & [5, 64, 8, 8, 8] & -- \\
$\vert\quad\vert-$Conv3d: 2-13 & [5, 64, 8, 8, 8] & 262,208 \\
$\vert\quad\vert-$InstanceNorm3d: 2-14 & [5, 64, 8, 8, 8] & 128 \\
$\vert\quad\vert-$LeakyReLU: 2-15 & [5, 64, 8, 8, 8] & -- \\
$\vert-$Sequential: 1-3 & [5, 64, 4, 4, 4] & -- \\
$\vert\quad\vert-$Conv3d: 2-16 & [5, 64, 4, 4, 4] & 262,208 \\
$\vert\quad\vert-$Instance
$\vert\quad\vert-$InstanceNorm3d: 2-17 & [5, 64, 4, 4, 4] & 128 \\
$\vert\quad\vert-$LeakyReLU: 2-18 & [5, 64, 4, 4, 4] & -- \\
$\vert-$Sequential: 1-4 & [5, 64, 2, 2, 2] & -- \\
$\vert\quad\vert-$Conv3d: 2-19 & [5, 64, 2, 2, 2] & 262,208 \\
$\vert\quad\vert-$InstanceNorm3d: 2-20 & [5, 64, 2, 2, 2] & 128 \\
$\vert\quad\vert-$LeakyReLU: 2-21 & [5, 64, 2, 2, 2] & -- \\
$\vert-$Sequential: 1-5 & [5, 1, 1, 1, 1] & -- \\
$\vert\quad\vert-$Conv3d: 2-22 & [5, 1, 1, 1, 1] & 4,097 \\ \midrule
\end{tabular}
\caption{Discriminator Architecture. A ModuleList consisting of 3 3D convolutional layers with progressively increasing number of filters. Each layer is followed by an InstanceNorm3D layer for normalization and a LeakyReLU activation function for non-linear activation. MaxPool3D layers are used after some Conv3D layers to downsample the feature maps. A series of Sequential blocks is consisted by another 4 conv3D layers, each containing an InstanceNorm3D layer, and a LeakyReLU activation function. These blocks help in progressively downsampling the feature maps, while keeping the number of channels (filters) constant. The final Sequential block contains a Conv3D layer that reduces the feature map dimensions to $1 \times 1 \times 1$, essentially providing the classification output. In total, the discriminator has 820,841 parameters.}
\label{appendixtab:discriminator_architecture}
\end{table}

\end{document}